\documentclass[preprint2]{aastex}
\def\gs{\mathrel{\raise0.35ex\hbox{$\scriptstyle >$}\kern-0.6em \lower0.40ex\hbo
x{{$\scriptstyle \sim$}}}}
\def\ls{\mathrel{\raise0.35ex\hbox{$\scriptstyle <$}\kern-0.6em \lower0.40ex\hbo
x{{$\scriptstyle \sim$}}}}

\shorttitle{Spectroscopic observation of galaxies}
\shortauthors{Carter et al.}

\begin{document}
 
\title{A photometric and spectroscopic study of dwarf and giant galaxies 
in the Coma cluster -  
V. Dependence of the spectroscopic properties on location in the cluster
\footnote{Based on observations made with the William Herschel Telescope 
operated on the island of La Palma by the Isaac Newton Group in the 
Spanish Observatorio del Roque de los Muchachos of the Instituto de 
Astrofisica de Canarias.}}

\author{
David Carter,$^{\!}$\altaffilmark{2}
Bahram\ Mobasher,$^{\!}$\altaffilmark{3,4}
Terry J.\ Bridges,$^{\!}$\altaffilmark{5}
Bianca M. Poggianti,$^{\!}$\altaffilmark{6}
Y. Komiyama,$^{\!}$\altaffilmark{7}
N. Kashikawa,$^{\!}$\altaffilmark{8}
M. Doi,$^{\!}$\altaffilmark{9} 
M. Iye,$^{\!}$\altaffilmark{8}
S. Okamura,$^{\!}$\altaffilmark{10,11} 
M. Sekiguchi,$^{\!}$\altaffilmark{12}
K. Shimasaku,$^{\!}$\altaffilmark{11}  
M. Yagi,$^{\!}$\altaffilmark{8}
N. Yasuda$^{\!}$\altaffilmark{8}
}
 
\smallskip

\affil{\scriptsize 2)  Astrophysics Research Institute, Liverpool John Moores University, Twelve Quays House, Egerton Wharf, Birkenhead, Wirral, CH41 1LD, UK}
\affil{\scriptsize 3) Space Telescope Science Institute, 3700 San Martin Drive, Baltimore, MD 21218, USA}
\affil{\scriptsize 4) Affiliated with the Space Sciences Department of the 
European Space Agency}
\affil{\scriptsize 5) Anglo-Australian Observatory, PO Box 296, Epping, NSW 2121, Australia}
\affil{\scriptsize 6) Osservatorio Astronomico di Padova, vicolo dell'Osservatorio 5, 35122 Padova, Italy}
\affil{\scriptsize 7) Subaru Telescope, 650 North Aohoku Place, Hilo, HI 96720, USA}
\affil{\scriptsize 8) National Astronomical Observatory, Mitaka, Tokyo, 181-8588, Japan}
\affil{\scriptsize 9) Institute of Astronomy, School of Science, University of
Tokyo, Mitaka, 181-0015, Japan}
\affil{\scriptsize 10) Research Center for the Early Universe, School of
Science, University of Tokyo, Tokyo 113-0033, Japan}
\affil{\scriptsize 11) Department of Astronomy, University of Tokyo,
Bunkyo-ku, Tokyo 113-0033, Japan}
\affil{\scriptsize 12) Institute for Cosmic Ray Research, University of Tokyo,
Kashiwa, Chiba 277-8582, Japan}

\begin{abstract}

We investigate the radial dependence of the spectroscopic properties, in 
particular the Mg$_2$, $<$Fe$>$ and H$\beta$ spectroscopic indices, in a sample
of galaxies spanning a wide range of absolute luminosity in the Coma cluster. 
After
allowing for the magnitude dependence of these indices, we find a significant 
gradient in Mg$_2$, in the sense that galaxies in the core of the cluster have
stronger Mg$_2$. We find only weak gradients in $<$Fe$>$ and H$\beta$. 
Using the model grids presented in an earlier paper in this series, we 
attribute the Mg$_2$ gradient to changes in metal abundance. One possible 
mechanism to create this abundance gradient is pressure confinement by the 
intracluster medium of material from Supernova driven winds early in the 
history of the galaxies.
\end{abstract}


\section{Introduction}

This is the fifth paper in a series which examines the properties of galaxies
in the Coma cluster in the magnitude range 12 $<$ R $<$ 20 and in a wide 
variety of environments. In this paper we examine the spatial dependence
of line indices which depend upon metal abundance and age, and of the 
luminosity weighted mean stellar abundances and ages which we derived from
these indices in Paper III of this series (Poggianti {\sl et al.} 2001). 
In this paper we examine the spatial dependence of the line indices
diagnostic of metal abundance and age and investigate the origin and
implication of such gradients.

The properties of galaxies in clusters have long been known to depend strongly
upon the environment. This was shown most clearly by the survey of Dressler
(1980a,b) who showed a dependence of galaxy morphology upon environment
in the sense that the proportion of E and S0 galaxies increases with increasing
density. This has been taken as evidence for a merger process occurring in 
clusters which transforms disk galaxies to spheroids at an early epoch. Further
work on clusters at higher redshifts (Dressler {\sl et al.} 1997) shows that
this dependence is qualitatively similar in highly concentrated clusters
at earlier epochs.

The importance of the intracluster medium has become clear more recently. 
the hot gas mass in clusters is typically 5 times the luminous mass in 
galaxies, and that gas typically has a metal abundance (specifically iron
abundance) of 0.2 - 0.5 solar (Finoguenov {\sl et al.} 2000; De Grandi
\& Molendi 2000). The majority of the heavy elements in a cluster are in the 
hot gas, but all of these heavy elements must have come from galaxies at
some stage. Finoguenov \& Ponman (1999) and Finoguenov {\sl et al} (2000)
study the distribution of both iron and $\alpha$ elements in the hot gas
component in clusters, and find that there are radial gradients in the 
abundances, stronger in iron than in $\alpha$ elements. They conclude that
Type II supernovae dominate the enrichment process in the outer regions of 
clusters, but in their centers Type I supernovae are more important. However
in the Coma cluster, which is the subject of this study,
Arnaud {\sl et al.} (2001) find that the abundance profile
is flat at 0.25 solar out to 15 arcminutes radius, except for a region of
enhanced abundance less than 30 arcsec in radius centerd on NGC 4874. These
authors present only a single abundance profile, presumably derived assuming 
solar abundance ratios.

As the galaxies are the source of the heavy elements, it is important to 
study and understand their properties. Line indices and luminosity weighted
ages and abundances for samples of Coma cluster galaxies have been published
by Jorgensen (1999) and Terlevich {\sl et al.} (2000), but our sample has
a much greater range in luminosity, and covers a wide range of radius and
of environment.

\section{Observations}

Our data are taken from photometric and spectroscopic observations of two
40 x 30 arcminute fields in the Coma cluster, with the Tokyo Mosaic CCD camera
and the WYFFOS fibre fed spectrograph respectively. Details of the 
observing setup and data reduction are given in earlier papers in this series.
Photometric observations and reduction are described in paper I (Komiyama
{\sl et al.} 2001). Paper II (Mobasher {\sl et al.} 2001) describes the 
spectroscopic sample selection. In paper III (Poggianti {\sl et al.} 2001) we 
derive line indices on the Lick system and luminosity weighted ages and 
metallicities from these indices. In the current paper we analyse
the following quantities: galaxy position, R magnitude, (B-R) color 
and concentration index from Paper I; and the spectroscopic indices Mg$_2$, 
$<$Fe$>$ and H$\beta$ from paper III, together with luminosity weighted ages 
and metallicities derived from the the model grids presented in the Mg$_2$
- H$\beta$ diagram in Figure 4 of that paper.

Our galaxy sample selection is described in paper II. The spectroscopic sample 
comprises two subsamples, a bright sample of galaxies at R$<$18 with known 
redshifts from Colless \& Dunn (1996) and Colless (1998) and a faint sample
selected from our own photometric imaging survey (Paper I) to have 
18$<$R$<$20 and 1$<$B--R$<$2. As described in Paper II, the samples are both 
believed to be free of morphological bias. Our faint spectroscopic sample is 
not complete, but the galaxies observed are drawn at random from the 
photometrically defined sample.

The sample discussed here comprises 229 galaxies from our spectroscopic
survey for which the formal errors on the Mg$_2$, $<$Fe$>$ and 
H$\beta$ indices are less than 0.04 magnitude, 1.5 {\AA} and 1.5 {\AA} 
respectively, and which did not show emission lines on visual inspection.
Of these, 163 are in the central Coma 1 field and 66 in the Coma 3 field 
which is located to the southwest of the cluster center.
described in Paper II. If more than one such spectrum was available then the 
line index values were combined with weighting proportional to the inverse of 
the variance on each line index. Because we require small errors in the
indices for inclusion in the sample we discriminate against 
faint galaxies and in particular low surface brightness galaxies, for low 
surface brightness the signal-to-noise ratio of our spectra is often too poor 
for inclusion. Rejecting emission lines (which otherwise will dilute the 
Balmer indices in particular) discriminates against galaxies with ongoing star
formation.

In this paper we examine the dependence of galaxy properties on the distance of
the galaxy from the the cluster center, which has been
taken to be at RA = 12h 59m 42.8s, Dec = 27$^{\circ}$ 58' 14'', Equinox 
J2000.0, a position near the peak of the X-ray emission (White {\sl et al.} 
1993) and some 1.8 arcminutes East of NGC 4874. However we do not completely 
sample the azimuthal distribution at radii greater than 15 arcminutes. Our 
outer field is to the South West of the center, and includes NGC 4839 and the 
group of galaxies around it, which are believed to be falling into the Coma 
cluster (e.g. Neumann {\sl et al.} 2001). The effect of the inclusion of this
group in our sample is considered later in the paper. 

\section{Radial dependence of line indices}

\begin{figure}
\plotone{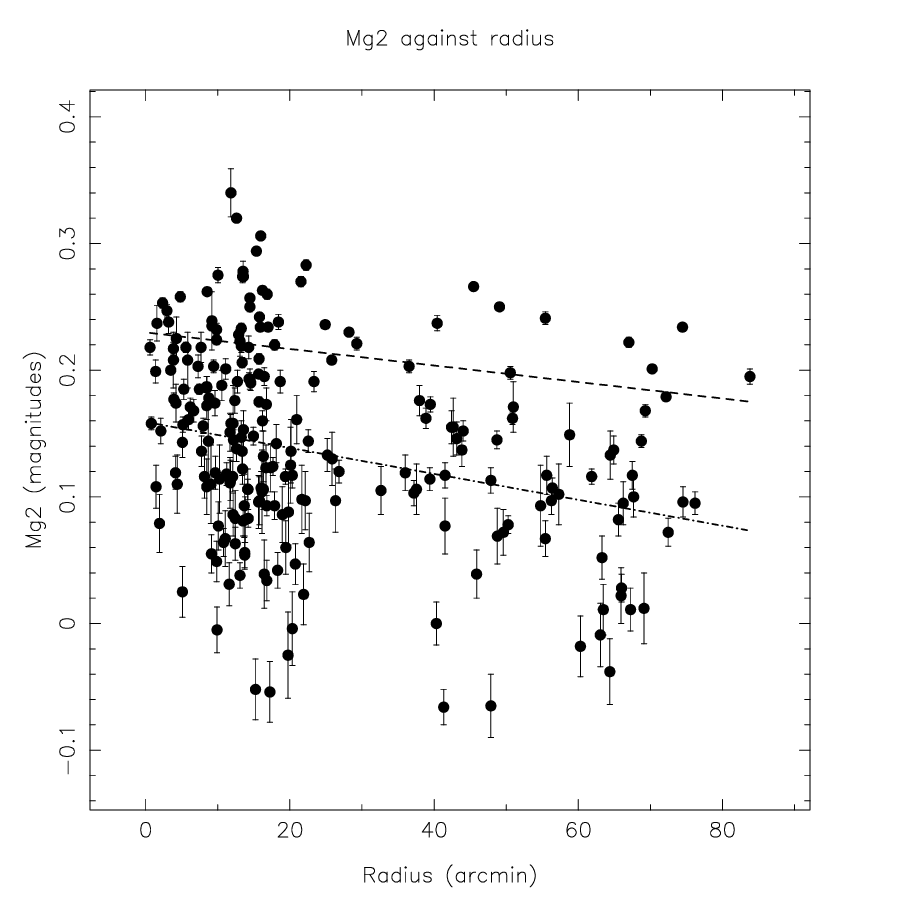}
\caption{Mg$_2$ index against radius for our sample of galaxies in the
Coma cluster. The dashed and dot-dash lines are weighted and unweighted
least squares straight line fits to the data respectively.
\label{Mgrad}}
\end{figure}

\begin{figure}
\plotone{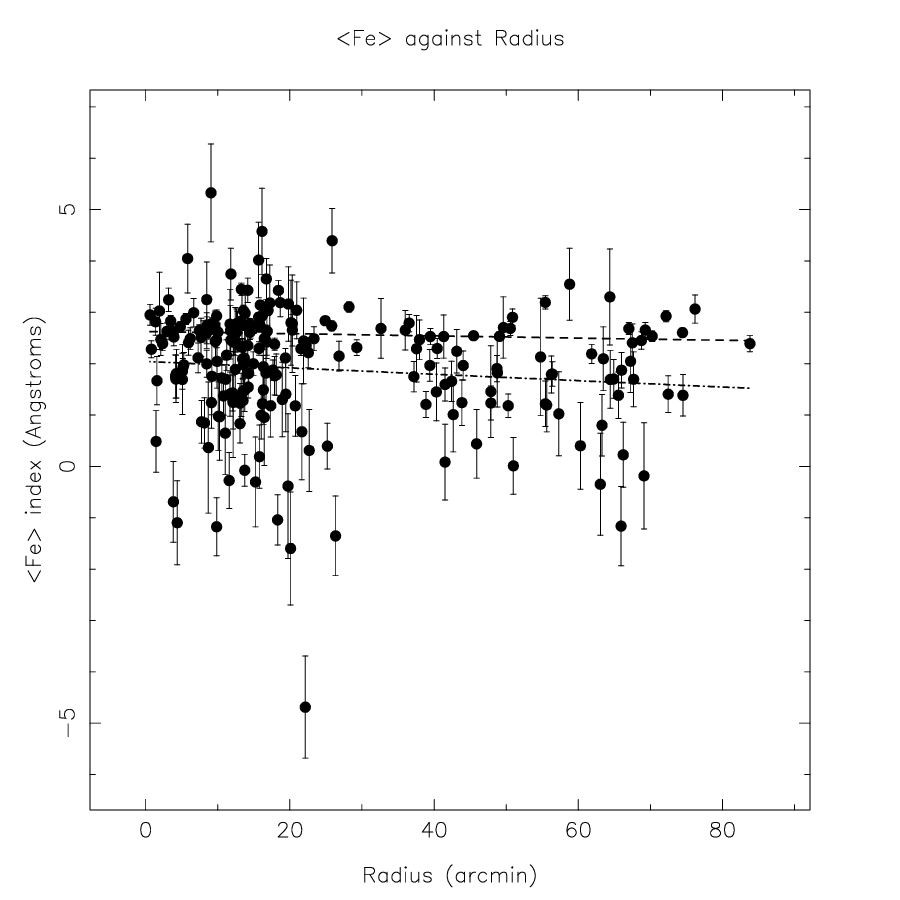}
\caption{$<$Fe$>$ index against radius for our sample of galaxies in the
Coma cluster. The dashed and dot-dash lines are weighted and unweighted
least squares straight line fits to the data respectively.
\label{Ferad}}
\end{figure}

\begin{figure}
\plotone{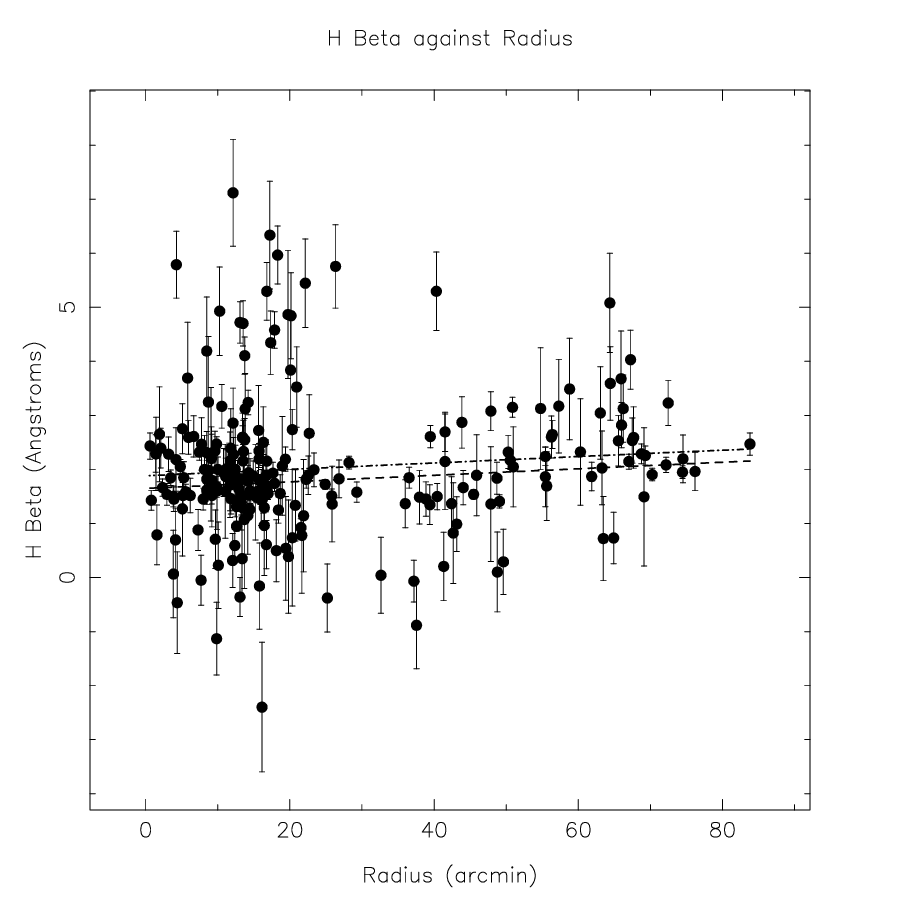}
\caption{H $\beta$ index against radius for our sample of galaxies in the
Coma cluster. The dashed and dot-dash lines are weighted and unweighted
least squares straight line fits to the data respectively.
\label{Hbrad}}
\end{figure}

In Figures \ref{Mgrad} - \ref{Hbrad} we plot the Lick Indices 
Mg$_2$, $<$Fe$>$ and H$\beta$ as a function of radius in the cluster. 

The Mg$_2$ index (Figure \ref{Mgrad}) shows a clear trend with radius, 
galaxies in the 
outer part of the cluster have lower Mg$_2$ indices than those in the core. 
There is a substantial scatter at all radii which is due to the strong trend
of this index with galaxy magnitude (paper III); the fainter galaxies have
weaker Mg$_2$, and larger errors, because of the lower signal-to-noise of
the spectra. The lines 
show weighted (dashed line) and unweighted (dot-dash line) least squares fits
to the data, there is a large offset between the lines because the brighter
galaxies which have stronger Mg$_2$ have better signal-to-noise and hence lower
errors and higher weights in the weighted fit. However both fits show the trend
which is visible in the data; towards lower Mg$_2$ at larger radii within
the cluster.

In Table 1 we present the statistical properties of the fits, to show the 
significance of the correlations. The important statistics in 
this context are ``Student's'' t, and P, the probability of obtaining a value 
of $|t|$ this high by chance in an uncorrelated sample.

The equivalent plots for the $<$Fe$>$ and H$\beta$ indices are shown in 
Figures \ref{Ferad} and \ref{Hbrad} respectively, the $<$Fe$>$ index 
shows a weaker gradient
than Mg$_2$, perhaps masked by higher errors, whereas H$\beta$ shows a marginal
positive gradient.

\begin{table*}
\begin{center}
{\small
\begin{tabular}{|l|l|l|l|l|l|l|l|l|l|l|}\hline
Fig&Ordinate&Abscissa&N&R&Slope&
$\epsilon_{Slope}$&Intercept&$\epsilon_{Intercept}$&t&
P(\%) \\ \hline
1&Mg$_2$&Radius&229&-0.22&-6.51x10$^{-4}$&1.91x10$^{-4}$&0.229&6.03x10$^{-3}$&
-3.41&0.08 \\
2&$<$Fe$>$&Radius&229&-0.08&-2.14x10$^{-3}$&1.69x10$^{-3}$&2.627&0.054&
-1.26&20.7 \\
3&H$\beta$&Radius&229&0.19&6.05x10$^{-3}$&2.03x10$^{-3}$&1.649&0.064&
2.99&0.31 \\
4a&Mg$_2$&R mag&229&-0.74&-3.58x10$^{-2}$&2.15x10$^{-3}$&0.7385&
3.16x10$^{-2}$&-16.7&$<10^{-3}$ \\
4b&$\delta$(Mg$_2$)&Radius&229&-0.25&-5.01x10$^{-4}$&1.27x10$^{-4}$&
1.21x10$^{-2}$&4.00x10$^{-3}$&-3.95&0.01 \\
5&$\delta$(B-R)&Radius&229&-0.45&-4.22x10$^{-3}$&5.63x10$^{-4}$&
0.1052&1.82x10$^{-2}$&-7.51&$<10^{-3}$ \\
6a&$<$Fe$>$&R mag&229&-0.56&-0.243&0.024&6.149&0.348&-10.3&$<10^{-3}$ \\
6b&$\delta$($<$Fe$>$)&Radius&229&-0.07&-1.37x10$^{-3}$&1.40x10$^{-3}$&
3.34x10$^{-2}$&4.45x10$^{-2}$&-0.98&33 \\
7a&H$\beta$&R mag&229&0.31&0.159&0.032&-0.541&0.475&4.94&$<10^{-3}$ \\
7b&$\delta$(H$\beta$)&Radius&229&0.19&5.45x10$^{-3}$&1.93x10$^{-3}$&
-0.133&0.061&2.82&0.5 \\
8&Metallicity&Radius&208&-0.16&-4.34x10$^{-3}$&1.89x10$^{-3}$&
-0.5028&0.0603&-2.30&2.2 \\
9a&Metallicity&R mag&208&-0.58&-0.1967&0.0193&2.619&0.319&-10.2&$<10^{-3}$ \\
9b&$\delta$(Metallicity)&Radius&208&-0.11&-2.46x10$^{-3}$&1.55x10$^{-3}$&
6.06x10$^{-2}$&4.95x10$^{-2}$&-1.59&11.4 \\
11&$\delta$(Metallicity)&Radius&143&-0.22&-4.20x10$^{-3}$&1.58x10$^{-3}$&
9.39x10$^{-2}$&4.55x10$^{-2}$&-2.66&8.8 \\
\hline
\end{tabular}}
\caption{Table showing the parameters describing the fits illustrated in 
Figures 1--9 and 11. Column 1 refers to the figure in the paper, columns 2 
and 3 list the dependent and independent variable respectively. Column 4 gives
the number of points used in each fit, and column 5 the correlation coefficient
between the variables. Column 6 gives the slope of the fitted straight line
(illustrated in the plot) and column 7 the formal error on this value. Columns
8 and 9 are the intercept with the vertical axis, and its error, respectively.
Column 10 gives ``Student's'' t statistic, which is related to the correlation
coefficient by $t = R \sqrt{{N-2}\over{1-R^2}}$, and is also identically equal 
to the ratio of the slope and its error. Column 11 gives the two-tailed 
(significance) probability, expressed as a percentage,
of a value of $|t|$ this high occurring by chance.}
\end{center}
\end{table*}

Although Figure \ref{Hbrad} shows marginal evidence for a positive gradient 
in H$\beta$ the statistical parameters in Table 1 show that this is not
significant. In this Figure we do not see
evidence for a concentration of Balmer line  strong galaxies in the 
NGC 4839 region of the Coma cluster, as was proposed by Caldwell {\sl et al.}
(1993), although our survey area does cover this region. NGC 4839 is 41.7 
arcminutes from our defined cluster center, and we see no concentration of 
H$\beta$ strong galaxies at this radius. Caldwell {\sl et al.}
measured H$\delta$ rather than H$\beta$, which is less affected by emission.
Emission line galaxies have been eliminated from our sample following visual
inspection, although some low level of emission may remain (possibly giving 
rise to the negative outliers in Figure \ref{Hbrad}). Again, these galaxies 
appear to be distributed evenly throughout the cluster. 

\subsection{Correcting for the luminosity dependence of the indices}

\begin{figure*}
\epsscale{2.0}
\plottwo{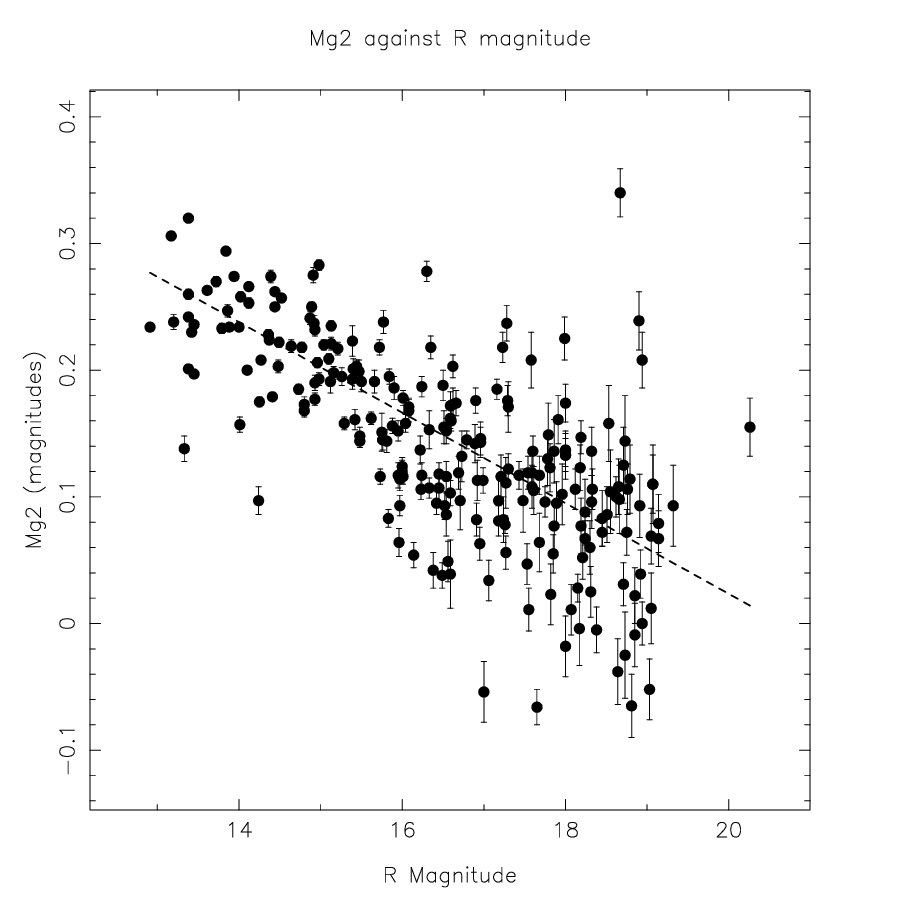}{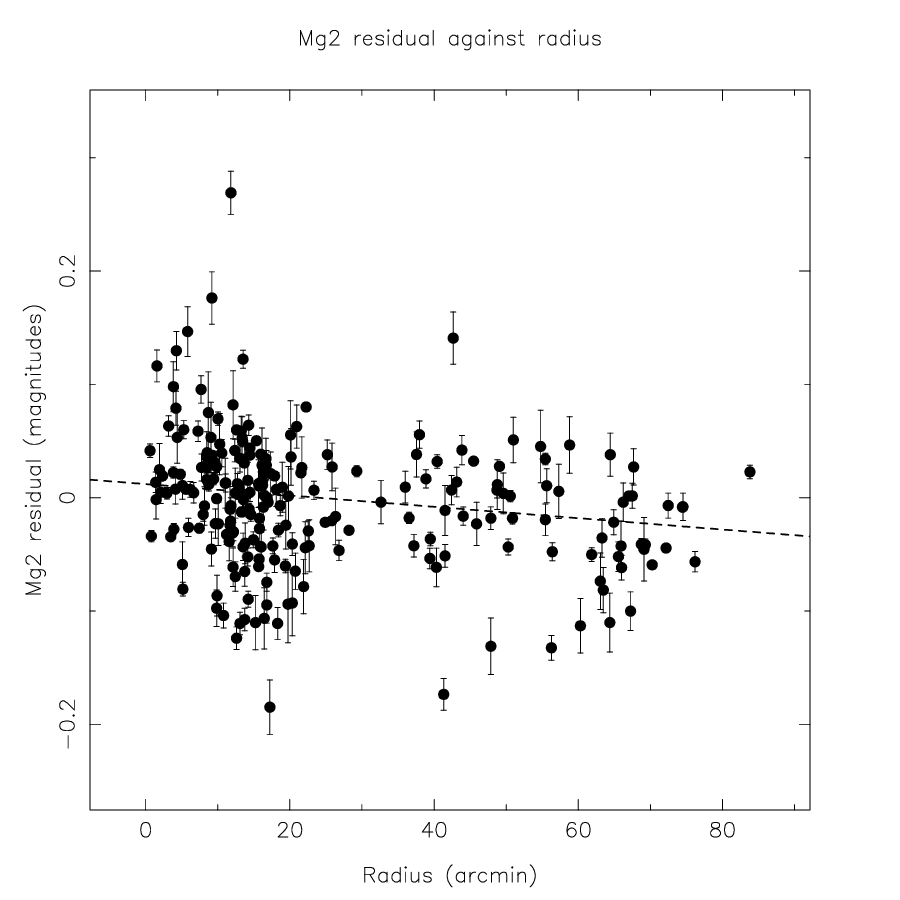}
\caption{a) Mg$_2$ against R magnitude for our sample, the dashed line is a 
weighted least squares straight line fit to the data. b) Mg$_2$ residual from 
the fit in (a) against radius 
within the cluster. The dashed line is a weighted least squares straight line 
fit. 
\label{Mgmag}
\label{Mgres}}
\end{figure*}

In Paper 3 we show a tight correlation between the 
metallicity derived from the Mg$_2$ index and galaxy luminosity for our 
sample. Here we 
investigate whether this dependence alone is sufficient to explain
the radial gradient
that we see in Figure \ref{Mgrad}. In Figure \ref{Mgmag}a we plot Mg$_2$ 
against R magnitude for 
the sample used here. This shows the well defined correlation between Mg$_2$
and magnitude, which is discussed in Paper 3. To take this 
relationship out of the radial dependence, in Figure \ref{Mgres}b we plot the 
residuals
in the Mg$_2$ index from a weighted least squares straight line fit against
radius. 

The correlation with radius is clearly still present in the residuals, and the 
scatter has been reduced, which 
shows that the correlation in Figure \ref{Mgrad} is not entirely due to 
index-magnitude relation.

\begin{figure}
\plotone{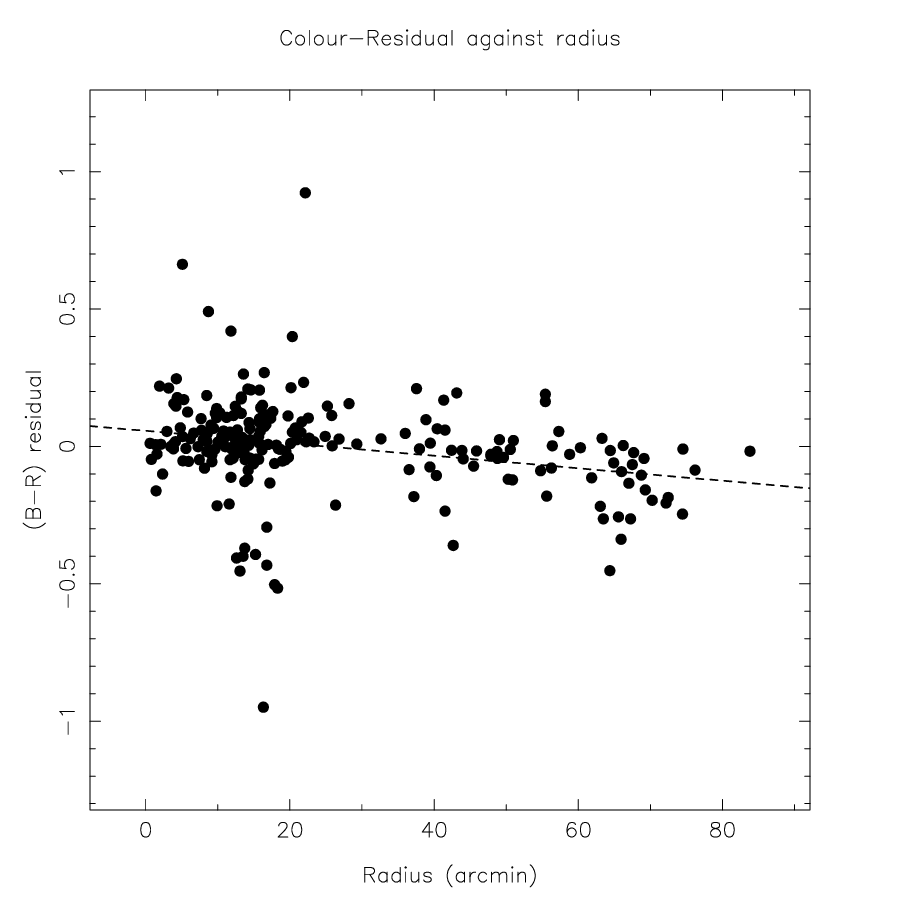}
\caption{Residual from the color-magnitude relation for the galaxies in the
spectroscopic sample, plotted against radius within the cluster. The dashed 
line is a least squares straight line fit to the data.}
\label{Colres}
\end{figure}

A similar effect is seen if we plot the residuals from the Color-Magnitude 
relation (paper I) against radius (Figure \ref{Colres}).
\epsscale{1.0}

\begin{figure*}
\epsscale{2.0}
\plottwo{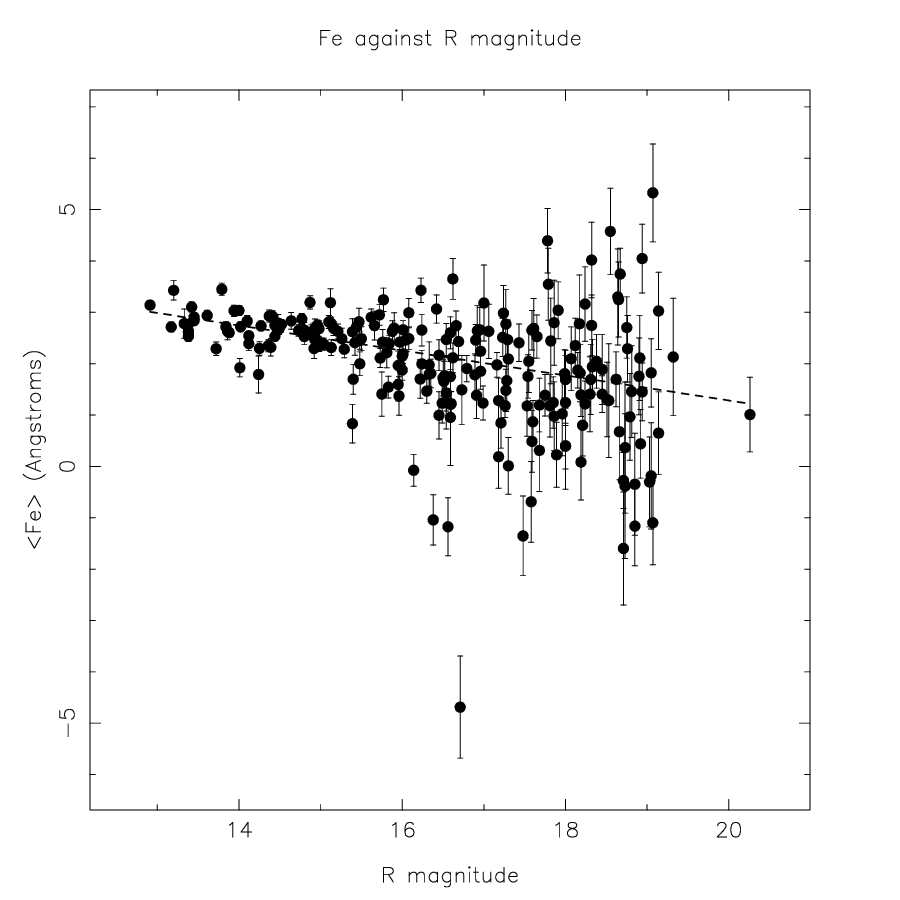}{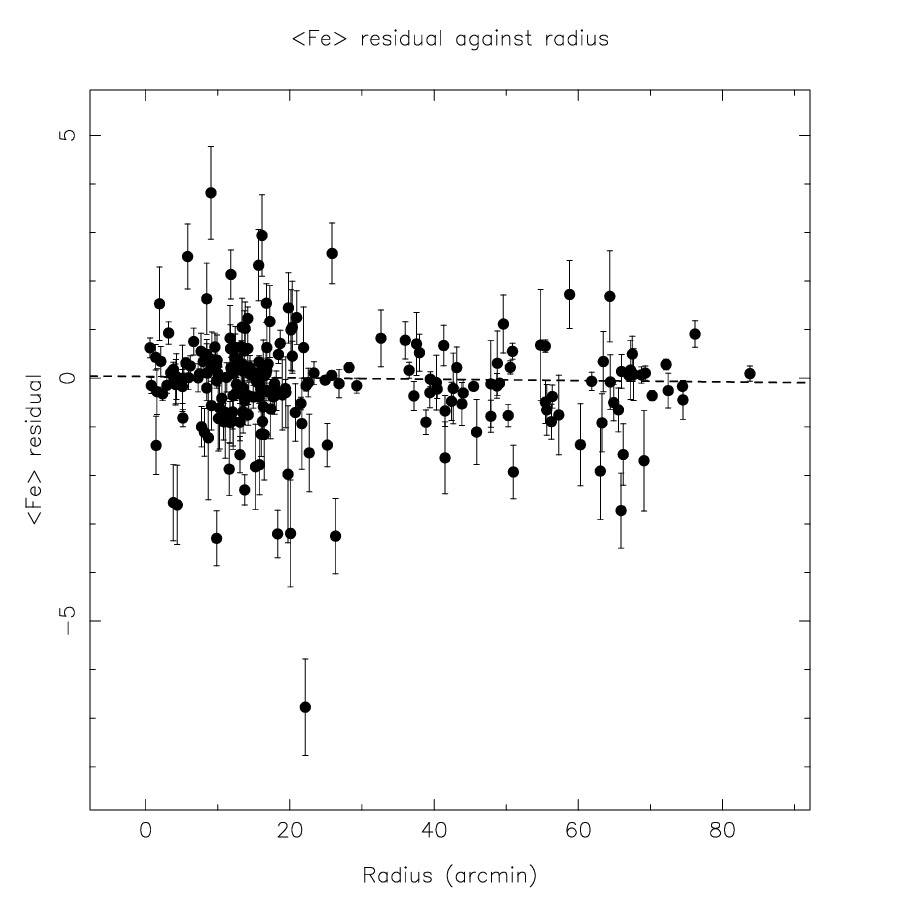}
\caption{a) $<$Fe$>$ against R magnitude for our sample. The dashed line is a 
weighted least squares straight line fit to the data. b) $<$Fe$>$ residual 
from the fit in (a) against radius within the cluster. The dashed line is a 
weighted least squares straight line fit.
\label{Femag}
\label{Feres}}
\end{figure*}

\begin{figure*}
\plottwo{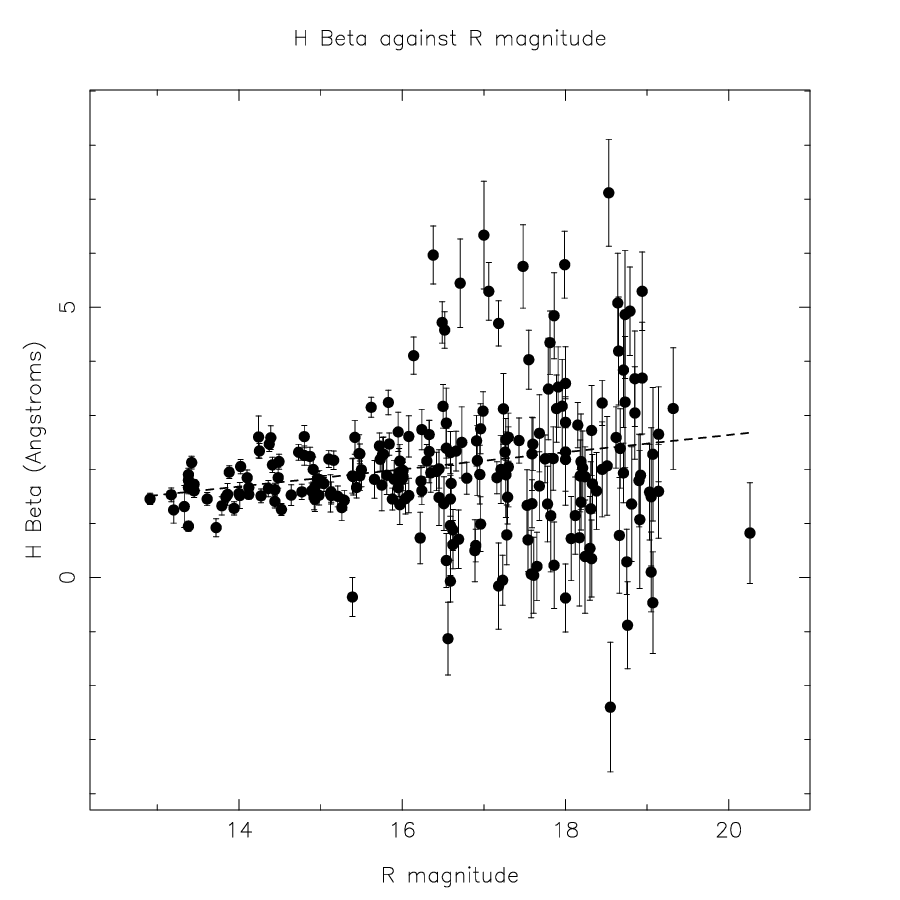}{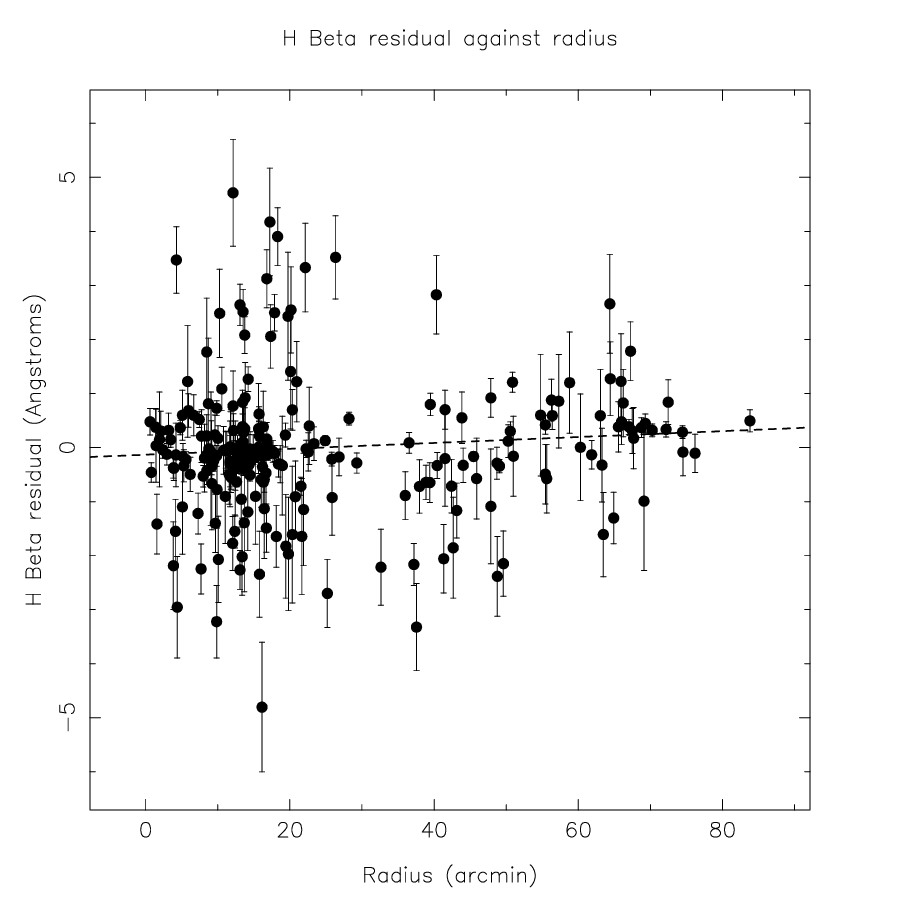}
\caption{a) H$\beta$ against R magnitude for our sample. The dashed line is a 
weighted least squares straight line fit to the data. b) H$\beta$ residual 
from the fit in (a), against radius within the cluster. The dashed line is a 
weighted least squares straight line fit. 
\label{Hbmag}
\label{Hbres}}
\end{figure*}

In Figures \ref{Femag} - \ref{Hbres} we repeat this procedure for our 
measurements of the $<$Fe$>$ and H$\beta$ indices, to see whether we can
identify residual radial trends in these indices which may have been masked by 
the scatter introduced by the index-magnitude relations. $<$Fe$>$ shows a 
strong dependence upon magnitude (Paper III and Figure \ref{Femag}a), and when
this relation is taken out we see no dependence of the residual upon radius
(Figure \ref{Feres}b). 

This result is surprising given the strong gradient
seen in Mg$_2$, and if confirmed suggests a different radial dependence of 
the enrichment mechanisms for these two elements. However we have to consider
the possibility that it is simply the larger scatter in $<$Fe$>$ which masks
a radial trend. Formally the slope in Figure \ref{Mgres}b is -0.00050 $\pm$ 
0.00013 mag arcmin$^{-1}$ and that in Figure \ref{Feres}b is -0.00137 $\pm$
0.00140 {\AA} arcmin$^{-1}$. If the slopes of the residual/radius relations
were to be in the same ratio of the slopes of the index/magnitude relations
for Mg$_2$ and $<$Fe$>$ (Figures \ref{Mgmag}a and \ref{Femag}a) then we would 
expect a slope in Figure \ref{Feres}b of -0.0034 {\AA} arcmin$^{-1}$, which is 
1.4$\sigma$ away from the observed value. 

\epsscale{1.0}

There is a very small positive radial gradient in the H$\beta$ residual, 
amounting to only 0.4 {\AA} over the entire radial range.

\subsection{Metallicity or age as a driver of the radial dependence of the
indices}

In paper 3 we interpreted our measurements of Lick indices in terms of 
luminosity weighted stellar age and metallicity with the help of Single
Stellar Population models described in that paper. The models show that
the Mg$_2$ index in particular is sensitive not only to the metallicity of
the population, but that at high metallicity it is also sensitive
to the age of the population. This problem is 
sometimes referred to as the ``age-metallicity degeneracy''. 
However, if galaxies really did consist of a single age 
and metallicity population, then a pair of indices such as Mg$_2$ and H$\beta$
would yield an unambiguous estimate of the age and metallicity of the 
population, subject to observational error and the correctness of a particular 
set of models such as those we use in Paper 3. 

The problems with interpretation of the Lick indices are largely due to the 
uncertainties in the models. First, the models are only Single Stellar 
Population models, and a galaxy such as our own does not consist of stars
with a single age and abundance. Second, our understanding of stellar evolution
is as yet incomplete. 
The limitations of current models are most apparent from our
lack of understanding of Galactic globular clusters. We 
still don't know if age is the dominant second parameter
affecting horizontal branch morphology (e.g. Chaboyer
et al. 1996; Stetson et al 1999; Rich et al. 1997), and
some authors claim that the age differences between
clusters of similar metallicity are too small to explain
their difference in HB morphology.  A related problem is
the embarrassing discrepancy between the spectroscopic
and CMD ages for 47 Tuc (Gibson et al. 1999), though
population synthesis models with improved physics reduce
the discrepancy (Vazdekis et al. 2001) or possibly eliminate
it completely (Schiavon et al. 2001).  Certainly, if we
cannot model the relatively simple globular clusters,
which are coeval and chemically homogeneous, we will have
little hope of understanding more complicated stellar 
systems such as galaxies.

Accepting these caveats, we use the models described in Paper III to 
investigate whether the radial gradients in the indices we observed are 
due to metallicity or age gradients, or a combination of the two. 

\begin{figure}[ht]
\plotone{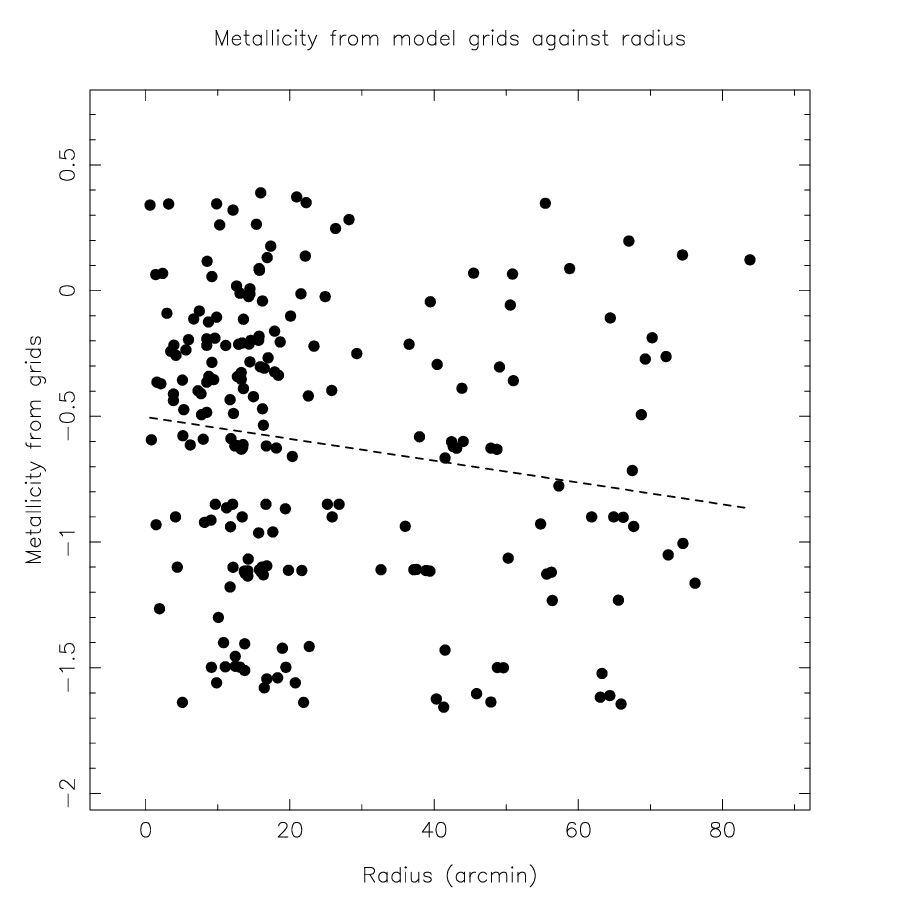}
\caption{Metallicity estimated from the model grid in the H$\beta$/Mg$_2$
diagram from Paper 3, plotted against radius within the cluster. As there
are now no formal errors on these values the dashed line represents an 
unweighted least squares fit.
\label{Metrad}}
\end{figure}

In Figure \ref{Metrad} we plot the metallicity from the grids in the 
H$\beta$/Mg$_2$ 
diagram in Paper 3 against radius, there is a gradient
of around 0.4 dex from the center of the cluster to 80 arcminutes radius.

\begin{figure*}
\epsscale{2.0}
\plottwo{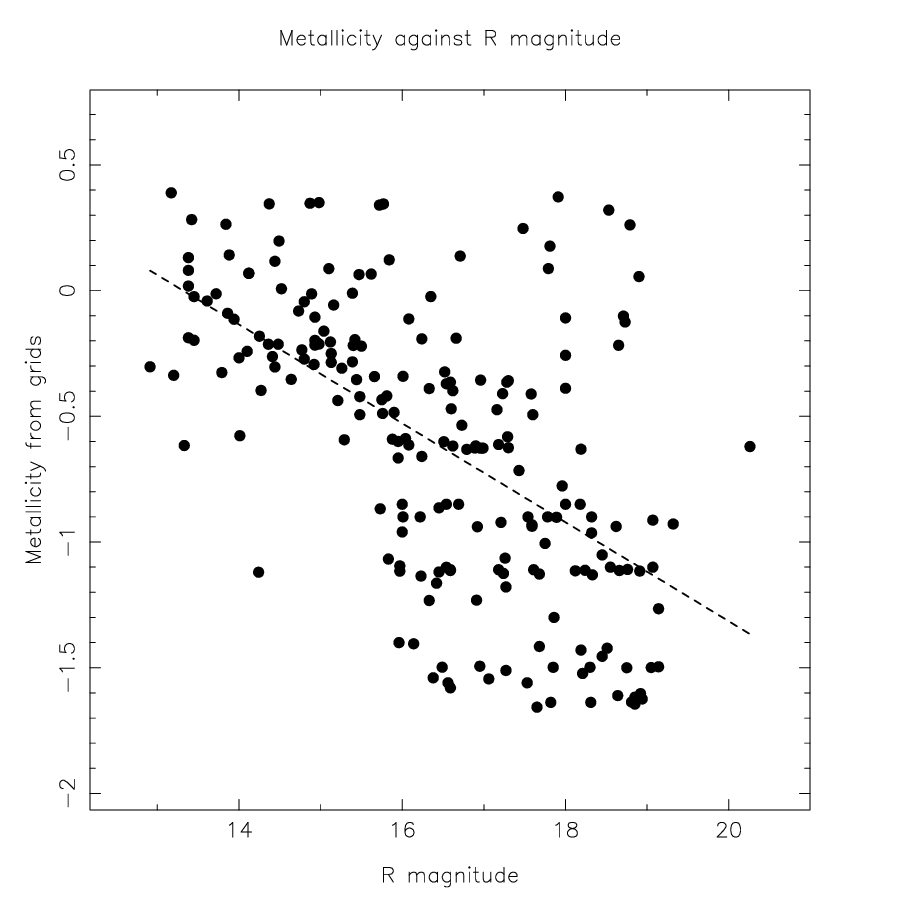}{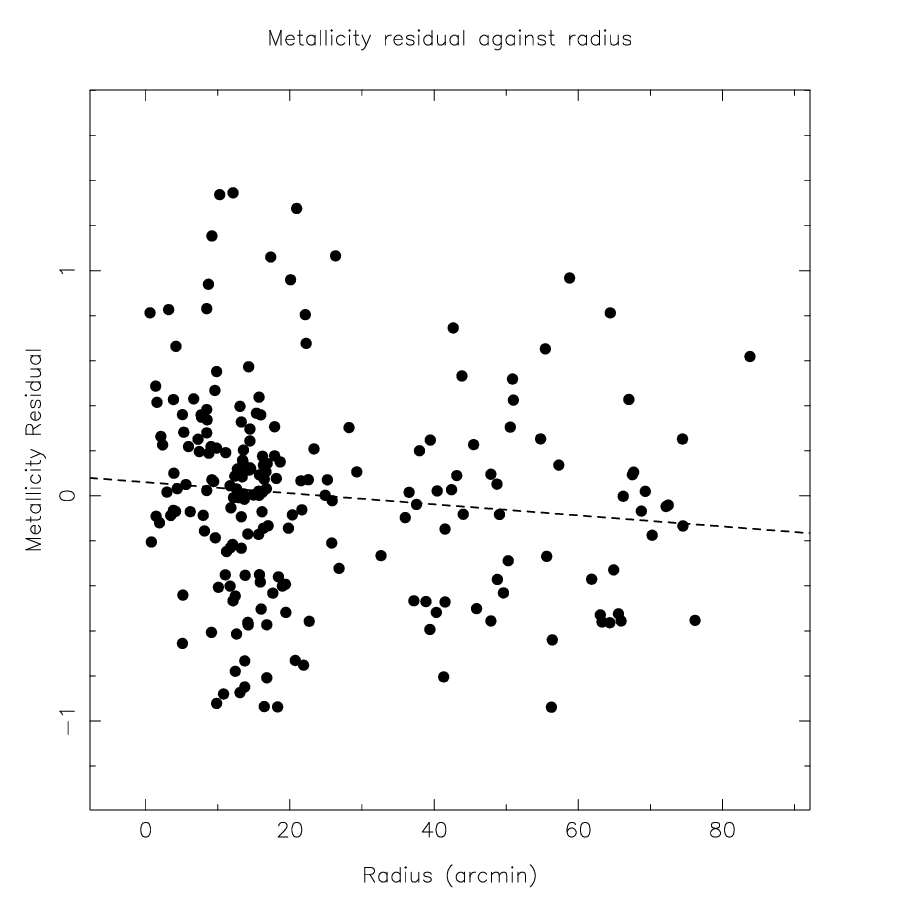}
\caption{a) Metallicity estimated from the model grid in the H$\beta$/Mg$_2$
diagram from Paper 3, plotted against R magnitude. The dashed line is an
unweighted least squares fit. b) Residuals from the fit in (a) plotted against 
radius within the cluster.
\label{Metmag}
\label{Metres}}
\end{figure*}
\epsscale{1.0}

In Figures \ref{Metmag}a and \ref{Metres}b we again take out the dependence 
of metallicity upon 
luminosity, Figure \ref{Metmag}a is the correlation between metallicity and R 
magnitude,
and in Figure \ref{Metres}b we plot the residuals from the fit to this 
relation against 
radius. When the strong luminosity dependence is taken out we find a gradient
$\Delta$log[Fe/H] $\sim$ -0.2 between the center of the cluster and 80 
arcminutes radius. The significance of this gradient is not as well established
as those in the raw indices, Table 1 shows an 11\% probability of a gradient 
of this size arising by chance, i.e. it is established at the 89\%
confidence level. Further observations of a larger sample of galaxies are 
required to firm up this result.  

A feature of Figure \ref{Metmag}a is the lack of galaxies in the lower left 
quadrant
of this plot, i.e. there are very few bright metal-poor galaxies, although
there are rather more faint metal-rich ones, although our spectroscopic
sample at R $<$ 16 is complete. This distribution, together
with the very strong correlation observed in this diagram, suggests that
processes which form smaller galaxies from larger ones (tidal stripping,
fragmentation and harassment) were more important in the evolution of the 
cluster than processes which form bright galaxies out of faint ones (e.g.
near equal-mass mergers), at least at times since the bulk of the
stars were formed.

\begin{figure}[ht]
\epsscale{0.8}
\plotone{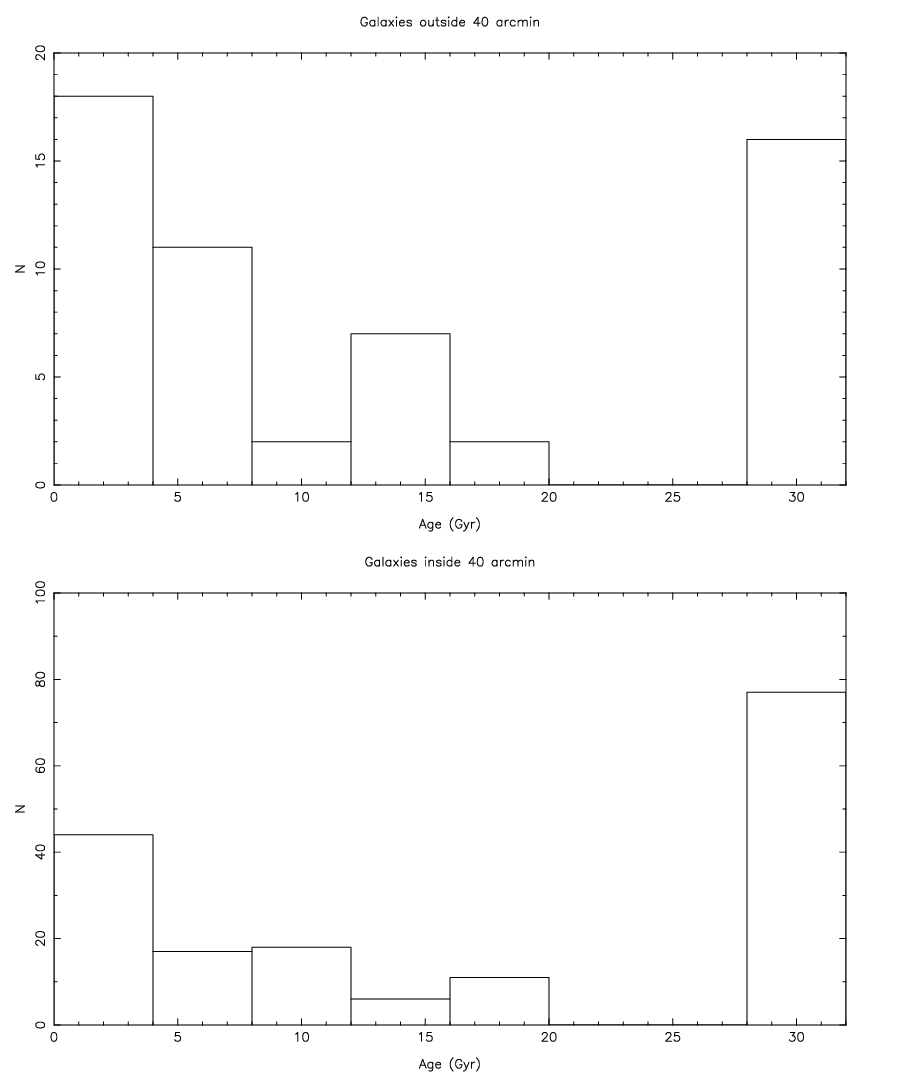}
\caption{Histogram of the luminosity weighted stellar age, for galaxies at 
radii greater than 40 arcminutes (upper panel) and less than 40 arcminutes
(lower panel). Galaxies which fall below the grid (i.e. at lower H$\beta$
strength than any models predict) in Figure 5 of Paper III are allocated
an age of 30 Gyr, which gives rise to the peaks at this age.  
\label{Agehist}}
\end{figure}

In Figure \ref{Agehist} we plot histograms of the derived luminosity weighted 
stellar age from the models in Paper III, divided into samples at projected 
radius greater than and less than 40 arcminutes. Our ages are derived from 
the right 
hand panel of Figure 5 of that paper. as pointed out there, many galaxies lie
below the grid lines, i.e. at lower H$\beta$ strength than any models predict,
these galaxies are allocated an age of 30 Gyr. The cause of this discrepancy is
discussed in Paper III. Figure \ref{Agehist} shows that a higher proportion
of the galaxies in the inner sample than in the outer lie below the grid. 
The ages attributed to these particular galaxies are not satisfactory for 
addressing the problem addressed in this paper, and because the 
significance of the difference in the age distributions outside and inside 40 
arcminutes depends critically upon our treatment of these galaxies, we can only
test such differences using non-parametric rank tests. Using the 
two-sample form of the Kolmogorov-Smirnov test, we test the 
samples against the null hypothesis that they are drawn from the same 
distribution. Taking at 
face value the rank of the ages of the galaxies in Figure \ref{Agehist}, 
and using the two-sample form of the test, we find that 
the distributions differ at the 93\% confidence level. However ignoring those 
galaxies placed at an age of 30 Gyr, we find that the age distributions for 
those galaxies with well established ages differ only at the 30\% confidence 
level, i.e. those distributions are entirely consistent with each other.

\epsscale{1.0}

Maraston \& Thomas (2000) present composite population models using isochrones
which reproduce the strong Balmer lines seen in old, metal-poor galactic 
globular clusters. Their models of early type galaxies consist of two old
populations, one metal poor and one metal rich. They propose that these models
can give rise to the observed scatter in cluster galaxy H$\beta$ strengths
as a combination of different proportions of old metal poor and old metal rich
populations without the need for a young population.  
If there were a metal poor population in the galaxies in our sample 
which became more important further out in the cluster, then this could mask
an age gradient, but this age gradient would be in the sense that galaxies 
at the center of the cluster were younger. There is no firm evidence for such 
an effect, and although the work of Maraston \& Thomas (2000) suggests that
there may be galaxies in our sample for which we are deriving 
unrealistically young ages, these galaxies seem to be spread evenly throughout
the cluster.

\subsection{Dependence of the abundance gradient on the morphology
and age}

It is well known that there is a dependence of galaxy morphology upon 
environment within clusters (Dressler 1980a,b), so in this section we consider 
whether the metallicity gradient which we see is the same effect. At first 
sight this seems unlikely, as the processes which give rise to transformations
between morphological types (ram pressure stripping, galaxy harassment) are
unlikely to affect the galaxy metallicity, and any effects on the 
metallicity-magnitude relation will result from effects on the luminosity, 
and is second order to effects upon the luminosity function. Moreover the 
processes which determine the galaxy metallicity are either
internal and depend primarily upon the galaxy mass (resulting in the 
metallicity-magnitude relation), or they are environmental and involve
transfer of gas between the galaxies and the intracluster medium, and are 
unlikely to depend upon morphology. 

Our sample consists largely of galaxies fainter than V=17, for which we have 
no objective or subjective morphological classification. Inevitably there
will be some spiral and irregular galaxies in the sample. Terlevich {\sl et 
al.} (2001) plot the (U-V) color-magnitude relation for galaxies of various
morphological types in the Coma cluster (their Figure 6). The color - 
magnitude (CM) relation is driven mostly by the metallicity-mass relation
(Faber 1973); Terlevich {\sl et al.} find that most spiral, irregular
and unclassified galaxies in their sample lie on the same CM relation as
the E and S0 galaxies, although there are a few outliers on the blue side of 
the ridge line. However these outliers are very blue, around 1 magnitude
bluer in (U-V) than the ridge line, which is likely to be due to ongoing
star formation. Such galaxies would have emission lines, and would have been 
rejected from our sample.

\begin{figure}[ht]
\plotone{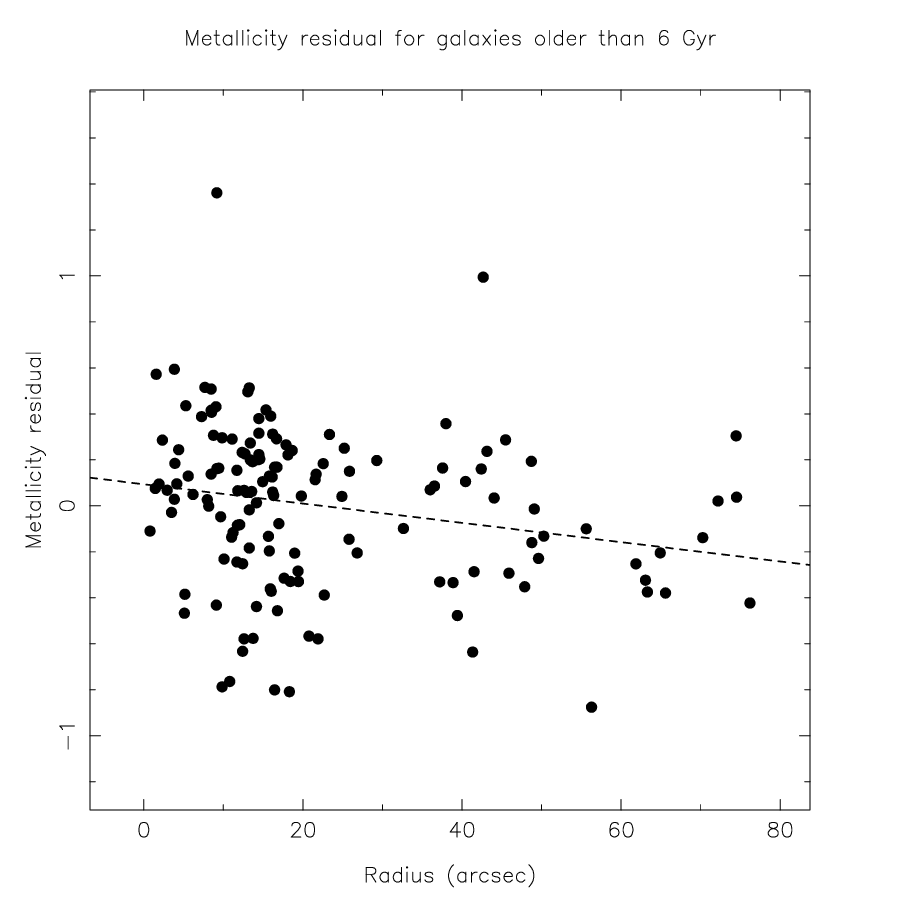}
\caption{Residuals from fit to the metallicity-magnitude diagram, 
plotted against radius within the cluster, only for those galaxies with mean
stellar age greater than 6Gyr.
\label{Metresold}}
\end{figure}

Figure \ref{Metresold} shows the residuals in metallicity from the metallicity
magnitude
relation, plotted only for those galaxies with a well established mean stellar
age greater than 6Gyr. There are only 143 galaxies in this plot, due to the
uncertainties in the ages. However the correlation with radius is still clear,
indeed it is stronger and more significant than in the complete sample 
(Figure \ref{Metres}b); Table 1, so the gradient is stronger in older 
galaxies, or at least galaxies with older stars.

\section{Discussion}

There is clear evidence for a metallicity gradient for the galaxies in the
Coma cluster, this presents itself as a difference of -0.03 magnitude in the 
Mg$_2$ index (Fig \ref{Mgres}b) or -0.2 in log[Fe/H] (Fig \ref{Metres}b) 
between the value
at the center of the cluster and that at 80 arcminutes radius. 

There are good reasons to expect that there might be a gradient in the 
metallicity of galaxies within clusters. Although galaxies will have 
undergone several orbits or crossings within the cluster they will not be
well mixed, so galaxies which form in the outer regions of a cluster will
spend most of their time there. The metallicity-magnitude relation can be 
understood in terms of the loss of processed material in supernova driven
outflows (Larson 1974, Vader 1984). This loss of the heavy elements is 
counteracted by the gravity of the galaxy, hence the metallicity-mass and 
metallicity-magnitude relation. But it can also be counteracted by external 
pressure of the intra-cluster medium (Silk {\sl et al.} 1987). Galaxies 
formed in the densest parts of the cluster would be expected to retain more 
of their heavy elements. We only detect a significant gradient in Mg$_2$,
not $<$Fe$>$, so it is possible that the mechanism which constrained the
processed material was more effective at a time when Supernovae Type II
were the dominant source of processed material, early in the history of
the cluster. This is also supported by the evidence of Figure 
\ref{Metresold}, that the gradient is stronger in galaxies with an older 
stellar component.

This picture is complicated by the possible modification of the cluster 
membership by accretion of subclusters. There is some evidence from the 
X-ray morphology (Neumann {\sl et al.} 2001) that the NGC 4839 group is 
being accreted  by the cluster. Any such accretion would be 
expected to dilute any gradient existing in the cluster. 

Guzm\'{a}n {\sl et al.} (1992), investigating the $\sigma$-Mg$_2$ relation
amongst the bright ellipticals in Coma, find an offset between a core sample 
and a halo sample in the sense that the halo ellipticals have weaker Mg$_2$
for a given $\sigma$. These authors, having only one index, cannot distinguish 
between metallicity and age as an origin of this offset, they show that age is
a possible origin. Rose {\sl et al.} (1994), using different indices, conclude
that there is a difference in stellar populations between galaxies in rich 
clusters and those in a field environment, in that the luminosity weighted
mean stellar population is younger and more metal rich in low density 
environments. They attribute this difference to 
an additional intermediate age population in the low density environments. They
do not investigate the difference between high- and low-density regions in 
the same cluster. We note that their dataset is heterogeneous, particularly
in the detectors used to gather it, and includes not only CCD spectra but 
also some from older image tube based detectors. It is not clear whether
the difference between their conclusions and ours results from different
observational techniques, or different sample selection, or because the
comparison between high- and low-density clusters is different from the 
comparison between regions of different density in the same cluster. 

Terlevich {\sl et al.} (2001) find a gradient in the (U-V)
color residual from the CM diagram with radius, in the sense that galaxies 
in the outer part of the Coma cluster are bluer than those in the center. At 
30 arcminutes radius, the galaxies are 0.03 magnitudes bluer in (U-V) than in 
the cluster center. Terlevich {\sl et al.} (2001) attribute this gradient to
a gradient in the mean stellar age of the galaxies, however (U-V) is also
sensitive to metallicity (Wallerstein 1962; Faber 1973). 
The relationship between (U-V) and metallicity in single stellar population 
models is complicated (e.g. Vazdekis {\sl et al.} 1996), and depends upon 
other parameters, and indeed differs by worrying amounts between models.
Inspection of Figure 3 of Vazdekis {\sl et al.} (1996) shows however that 
a gradient of -0.1 in log[Fe/H] between 0 and 30 arcminutes radius would 
be sufficient to produce the gradient in (U-V) seen by Terlevich 
{\sl et al.} without the need for an age gradient.

Balogh {\sl et. al} (2000) propose a model in which star forming galaxies
fall into rich clusters at a fairly uniform rate over time, and their star 
formation is gradually quenched by ram pressure and tidal stripping removing
their gas over a period of a few Gyr. Their model explains the observed 
gradients in star formation rate, as measured by spectroscopic indices
in the CNOC cluster sample (Balogh {\sl et al.} 1999). This mechanism would
be expected to lead to an age gradient in our sample, indeed given the small
sample and the uncertainties in the ages of a substantial number of galaxies in
Figure \ref{Agehist} we cannot rule this out. However as galaxies further out
in the cluster would have had their star formation truncated on average later
than galaxies in the core, any effect on the metallicity gradient would be 
in the opposite sense to that which we see. We conclude that the metallicity 
gradient is not due to infall of galaxies. 

\section{Conclusions}

We detect significant evidence for a negative radial gradient in the 
metallicity of the galaxies in the Coma cluster, as measured from pairs of 
indices including Mg$_2$. Our $<$Fe$>$ index measurements are less accurate, 
however the gradient that we find in this index is significantly shallower
that we would deduce from the size of the Mg$_2$ gradient. 
Evidence for a gradient in luminosity weighted stellar age is unclear, and we
suggest that the gradient in (U-V) observed by Terlevich {\sl et al.} (2001)
and attributed by them to an age gradient could instead be driven by the 
metallicity gradient that we observe.

Our observations are most easily explained by models in which the pressure
of the intracluster medium confines enriched material in supernova driven
winds early in the history of the cluster. We find no firm evidence for a 
gradient in the mean stellar ages of galaxies with radius. Thus our 
observations support models in which the epoch of formation of the
bulk of the stars in galaxies does not depend upon their location
within the cluster.

\acknowledgments

\pagebreak

\end{document}